\documentclass[conference]{IEEEtran}

\usepackage{cite} 
\usepackage{graphicx} 
\usepackage{amsmath} 
\usepackage{amsfonts}
\usepackage{algorithm} 
\usepackage{algpseudocode} 
\usepackage{array} 
\usepackage{subfig}
\usepackage{color}

\usepackage{url} 
\usepackage{mathtools}

\DeclarePairedDelimiter\abs{\lvert}{\rvert}%
\begin{document}
\title{Hardware-Based ADMM-LP Decoding}

\author{\IEEEauthorblockN{Mitchell Wasson \ Mario Milicevic \ Stark C. Draper \ Glenn Gulak}
\IEEEauthorblockA{University of Toronto\\
	Email: m.wasson@mail.utoronto.ca \ mario.milicevic@utoronto.ca \ stark.draper@utoronto.ca \ gulak@eecg.toronto.edu}
}

\maketitle

\begin{abstract}
In this paper we present an FPGA-based implementation of
linear programming (LP) decoding. LP decoding frames error correction
as an optimization problem. This is in contrast to variants of belief
propagation (BP) decoding that view error correction as a problem of
graphical inference. There are many advantages to taking the
optimization perspective: convergence guarantees, improved
performance in certain regimes, and a methodology for incorporating
the latest developments in optimization techniques. However, LP
decoding, when implemented with standard LP solvers, does not easily
scale to the blocklengths of modern error-correction codes. In
earlier work, we showed that by drawing on decomposition methods from
optimization theory, specifically the alternating direction method of
multipliers (ADMM), we could build an LP decoding solver that was
competitive with BP, both in terms of performance and speed. We also
observed empirically that LP decoders have much better high-SNR
performance in the ``error floor'' regime, a trait of particular
relevance to optical transport and storage applications. While our
previous implementation was in floating point, in this paper we report
initial results of a fixed-point, hardware-based realization of our
ADMM-LP decoder.
\end{abstract}

\section{Introduction}
The field of error-correction coding was revolutionized in the
mid-1990s by the widespread adoption (and academic study) of
graph-based codes and associated message-passing decoding algorithms.
A key aspect of the success of these codes was their compatibility
with hardware. BP-based decoders are naturally
distributed algorithms and variants such as Min-Sum are (relatively)
easily mapped to hardware. Graph-based codes, particularly turbo
codes and low-density parity-check (LDPC) codes, have been adopted
in many real world systems.

In the early 2000s, Feldman and his
collaborators realized that the maximum
likelihood (ML) decoding problem for binary linear codes can be
rephrased as an integer program~\cite{feldman_2005_journal}. One
obtains an LP by relaxing the integer constraints.  
Feldman's results generated much interest among
coding theorists. LPs are an extremely well-studied and understood
class of optimization problems, especially when contrasted with BP.
For instance, LP decoding has an ML certificate
property~\cite{feldman_2005_journal}. If LP decoding fails, it fails in a
detectable way (to a non-integer vertex) and the relaxation can be
tightened and the LP re-run~\cite{seigel_2008_adaptive_lp}. If a high-quality
expander or high-girth code is used, LP decoding is guaranteed to
correct a constant number of bit flips~\cite{feldman_2005_constant,arora_2009_constant}.
Broadly, it was hoped that by studying LP decoding, more would be
understood about BP decoding.

On the practical side, there was less excitement. There initially
seemed to be no real-world need for such a decoder, and further,
traditional LP solvers did not scale easily to the
blocklengths of modern error-correcting codes. Nevertheless, a number of
groups did study how to build an application-specific low-complexity LP decoder~\cite{vontobel_2006_low_complex_lp,burshtein_2009_iterative_lp,seigel_2008_adaptive_lp,draper_2013_admm_lp}.
In particular, Barman et al. built an
application-specific LP decoder that was computationally competitive with BP and that had a message-passing structure with a standard message schedule \cite{draper_2013_admm_lp}. They solved the LP decoding
problem using ADMM,
a decomposition technique used in large-scale optimization. Able to
study LP decoding performance at long blocklengths, it was observed
empirically, and later confirmed
theoretically that in the
high-SNR regime LP decoders far outperform BP \cite{draper_2013_admm_lp,liu_2014_instanton,liu_2015_jumpLinear}. In this regime, BP decoders often experience an ``error floor'' while LP decoders do not.
Further, LP decoding can be used as a subroutine in a multi-stage
decoder that quickly approaches ML performance~\cite{wang_2009_multistage}.
Thus, for application areas in which reliability demands are extreme, LP decoding is an
attractive alternative or complement to BP.

However, one major hurdle remains that will determine whether or
not ADMM is truly a viable competitor to BP in high-reliability
applications. That hurdle is to show that ADMM-LP decoding
algorithms can be mapped to hardware without unacceptable performance loss.

In this paper, we present an FPGA-based implementation of an ADMM-LP
decoder. First, we review recent developments made to implement the key
computational primitive that underlies ADMM in hardware~\cite{wasson_2015_pp}. This primitive is 
a Euclidean projection onto a particular convex object termed the ``parity polytope.''
Then we describe how to assemble the pieces to form a complete LP decoder. We
present results for a [155, 64, 20] quasi-cyclic (QC) LDPC code introduced by Tanner et
al.~\cite{tanner_2001_refcode}, as well as the [672, 546] QC-LDPC code in the IEEE 802.11ad (WiGig) standard~\cite{IEEE-80211ad}. We test code performance using a full FPGA-based simulation environment. Our initial investigation reveals ADMM-LP decoding in hardware requires more resources than BP-based decoders. However, we also find that it is possible to achieve competitive error rate results with such a fixed-point implementation.

\section{Background}
\subsection{LP Decoding}
In this paper, we consider the decoding of binary linear codes. A binary linear code $\mathcal{C}$ can be defined by an $m\times n$ parity-check matrix $H$ as $\mathcal{C} = \left\{ x \in \{0,1\}^n : Hx = 0 \pmod{2} \right\}$. Each parity-check matrix row corresponds to a check, which specifies a subset of bits that must add to 0 modulo 2. These checks are indexed by the set $\mathcal{J} = \{1, \dots, m\}$. Each column of the parity-check matrix corresponds to a codeword symbol or variable indexed by $\mathcal{I} = \{1, \dots, n\}$. The neighborhood of check $j$, denoted $\mathcal{N}_c\left(j\right)$, is the set of variables that check $j$ says must add to 0. That is, $\mathcal{N}_c\left(j\right) = \{ i : H_{j,i} = 1 \}$. Similarly, the neighborhood of variable $i$, denoted $\mathcal{N}_v\left(i\right)$, is the set of checks that variable $i$ participates in.

It was shown that maximum likelihood (ML) decoding of binary linear codes over symmetric memoryless channels is equivalent to the minimization of a linear objective function~\cite{feldman_2003_phd,feldman_2005_journal}. The linear objective is formed by creating the vector of log-likelihood ratios $\gamma$, where $\gamma_i = \log \left( \frac{p(y_i|x_i=0)}{p(y_i|x_i=1)} \right)$. Here $y_i$ denotes the $i^{\text{th}}$ received channel output symbol, and $x_i$ is $i^{\text{th}}$ transmitted codeword symbol. The resulting ML decoding problem is $\arg\min_{x\in \mathcal{C}} \gamma^\top x$. Note that $\gamma$ can be multiplied by any positive scalar without changing the decoding problem.

\begin{figure*}
	\centering
	\subfloat[Identify the active facet]{
		\label{ppPic}
		\includegraphics[width=.25\textwidth]{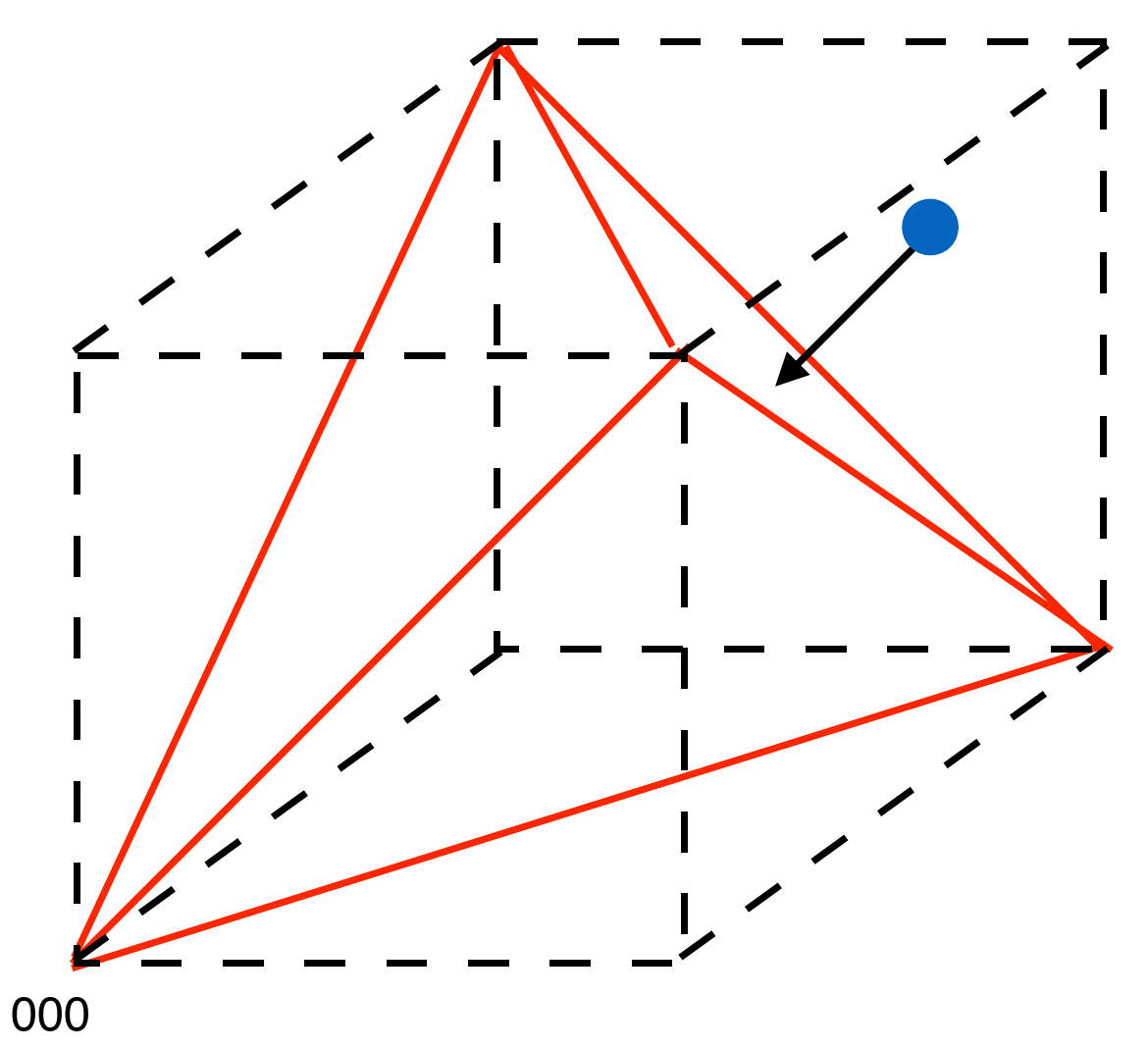}} \hfill
	\subfloat[Transform to reorient problem]{\includegraphics[width=.25\textwidth]{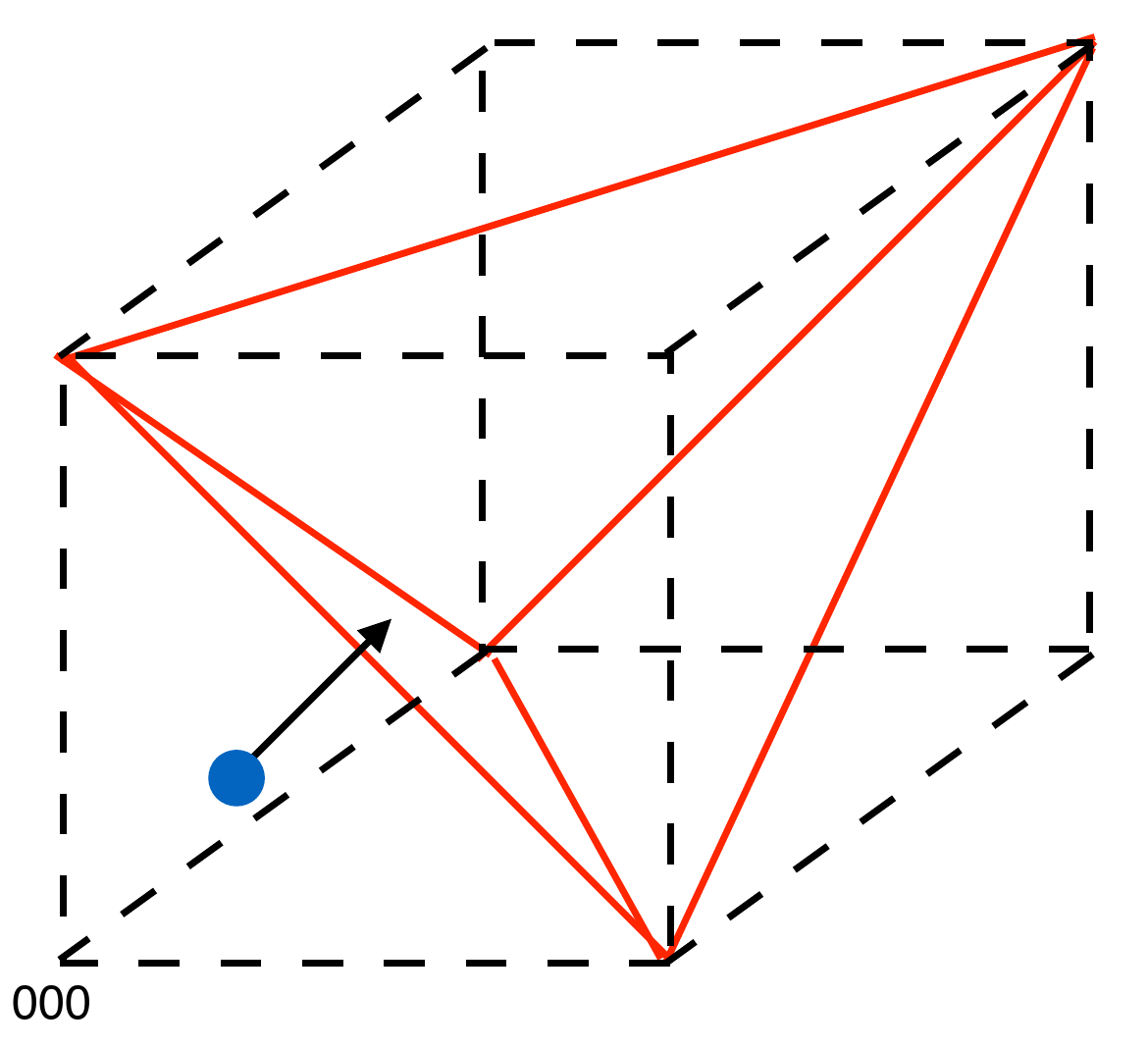}} \hfill
	\subfloat[Project onto probability simplex]{\includegraphics[width=.25\textwidth]{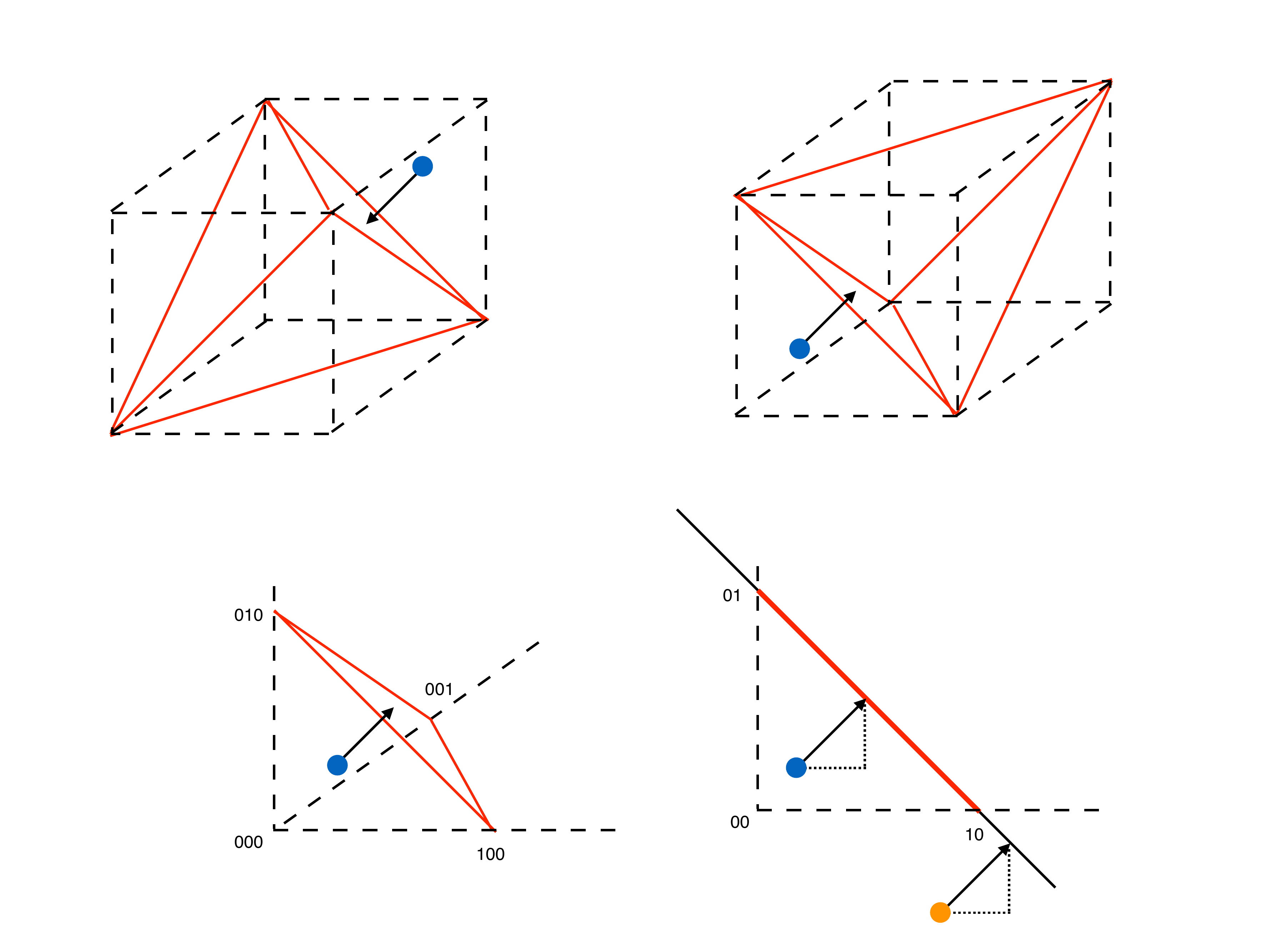}} \hfill
	\caption{Projection onto the parity polytope $\mathbb{PP}_3$: Identify the active facet, similarity transformation, simplex projection.}
	\label{ppProject}
\end{figure*}

Let $x_{S}$, $S\subseteq \mathcal{I}$ be the length $\abs{S}$ vector formed with the components of $x$ indexed by $S$. With this notation, we can restate the parity-check condition for a valid codeword as $\mathcal{C} = \left\{ x \in \{0,1\}^n : 1^\top x_{\mathcal{N}_c\left(j\right)} = 0 \pmod{2} \text{ for all } j\in\mathcal{J} \right\}$. This states that codeword variables connected to a check must be an even-weight vertex of the unit hyper-cube. Linear program (LP) decoding results from relaxing such constraints \cite{feldman_2003_phd, feldman_2005_journal}. LP decoding requires codeword variables connected to a check be in the convex hull of the even-weight vertices of the unit hyper-cube. Visualized in Fig.~\ref{ppPic}, the convex hull of the even-weight vertices of the unit hyper-cube is referred to as the parity polytope. The formal definition of the $d$-dimensional parity polytope, denoted $\mathbb{PP}_d$, is
\begin{equation*}
\mathbb{PP}_d := \text{conv}\left(\left\{ e \in \left\{ 0,1 \right\}^d
: 1^\top e = 0 \pmod{2} \right\}\right).
\end{equation*}

With the parity polytope relaxation, LP decoding is the following optimization problem:
\begin{equation}
\label{LPDecoding}
{\begin{array}{rll}
		\min\limits_{x} & \gamma^\top x\\
		\text{subject to}& x_{\mathcal{N}_c\left(j\right)} \in \mathbb{PP}_{\left\vert{\mathcal{N}_c(j)}\right\vert}  & j\in\mathcal{J}
	\end{array}}
\end{equation}

\subsection{ADMM Decomposition}

More recently, a distributed message-passing algorithm to accomplish LP decoding was developed~\cite{draper_2013_admm_lp}. This algorithm was created by applying the ADMM decomposition technique to LP decoding~\cite{boyd_2011_admm}. The decomposition starts by adding $\abs{\mathcal{J}}$, auxiliary ``replica'' variable vectors, $z_j$, for all $j \in \mathcal{J}$. These $z_j$'s are length $\mathcal{N}_c(j)$ vectors that correspond to the codeword variables participating in check $j$. Additionally, the constraint that $x$ must be in the unit hyper-cube is added. The result is the equivalent LP:

\begin{equation}
\label{ADMMLPDecoding}
{\begin{array}{rll}
		\min\limits_{x, z} & \gamma^\top x\\
		\text{subject to}	&z_j = x_{\mathcal{N}_c\left(j\right)} & j\in\mathcal{J}\\
		& z_j \in \mathbb{PP}_{\left\vert{\mathcal{N}_c(j)}\right\vert}  & j\in\mathcal{J}\\
		& x \in \left[0,1\right]^n 
	\end{array}}
\end{equation}
where $z$ aggregately refers to the $z_j$'s. The decoding algorithm that results from the application of ADMM to (\ref{ADMMLPDecoding}) is an iterative updating of $x$ and the $z_j$'s. After each iteration, an update of the dual variable vectors, $\lambda_j$, also occurs. The development of this algorithm is presented in \cite{draper_2013_admm_lp}. We present a slightly modified version of the original algorithm to make the message-passing structure more explicit.

\begin{algorithm}[h]
	\caption{ADMM-LP Decoding Algorithm}
	\label{decodingALgorithm}
	
	\textbf{Input:}  LLR vector $\gamma \in \mathbb{R}^n$ and iteration cap $B$\\
	\textbf{Output:} Decoding $x\in [0,1]^n$
	\begin{algorithmic}[1]
		\State $b=0$
		
		\ForAll{$j\in \mathcal{J}$}
		\State $\lambda_j = 0$
		\ForAll{$i\in \mathcal{I}$}
		\State $m_{j \rightarrow i} = \frac{1}{2}$ 
		\EndFor
		\EndFor
		
		\While{$b<B$}
		
		\ForAll{$i\in \mathcal{I}$}
		\State $x_{i} = \prod\nolimits_{\left[ 0,1 \right]}\left( \frac{1}{\abs{\mathcal{N}_v\left( i \right)}} \label{step.varUpdate}
		\left( 1^\top m_{\mathcal{N}_v\left( i \right) \rightarrow i}  -\gamma_i\right) \right)$
		\EndFor
		
		
		\ForAll{$j\in \mathcal{J}$}
		
		
		\State $v = x_{\mathcal{N}_c\left( j \right)} + \lambda_j$
		
		\State $z = \prod\nolimits_{\mathbb{PP}_{\abs{\mathcal{N}_c\left( j \right)}}}\left( v \right)$ \label{step.chkUpdate}
		\State $\lambda_j = v - z$
		
		\State $m_{j \rightarrow \mathcal{N}_c\left( j \right)} = 2z - v$
		\EndFor
		\State $b=b+1$
		\EndWhile
		
		\State \textbf{return} $x$
		
	\end{algorithmic}
\end{algorithm}

In Algorithm~\ref{decodingALgorithm}, we use the notation $m_{\mathcal{N}_v\left( i \right) \rightarrow i}$ to refer to the length $\abs{\mathcal{N}_v\left( i \right)}$ vector whose components are the messages sent to variable $i$ from its neighbors. Similarly, $m_{j \rightarrow \mathcal{N}_c\left( j \right)}$ refers to the length $\abs{\mathcal{N}_c\left( j \right)}$ vector whose components are the messages from check $j$ to its neighbors.

This presentation of the decoding algorithm shows that messages are passed between variable and check computations in a manner similar to BP-based decoding. There are two differences in the message-passing structure though. The first is that variable $i$ sends the same message $x_i$ to all its neighbors. The second is the addition of the dual variable vectors $\lambda_j$. These vectors serve as an internal state for checks.

We also have an intuition for the operations performed in the variable and check updates. The variable update (step~\ref{step.varUpdate}) is an averaging operation of all incoming messages along with the channel information. The average is then projected onto the feasible set for $x$. Check computations correspond mainly to a projection onto the parity polytope to enforce code constraints (step~\ref{step.chkUpdate}). The $\lambda_j$'s are incorporated into these projections to achieve faster convergence by indicating how the check has been violated in the past.

Projecting onto the parity polytope is the
essential nontrivial primitive in ADMM-LP decoding. We
review this projection and overview an algorithm that accomplishes it. 

The parity polytope is a polyhedron. Therefore, projecting a point $v \in \mathbb{R}^d$ onto
$\mathbb{PP}_d$ is given by a quadratic program,
\begin{equation*}
	\prod\nolimits_{\mathbb{PP}_d}\left( v \right) := \arg\min_{z \in \mathbb{PP}_d} \| v - z \|_2^2.
\end{equation*}

Barman et al. developed a routine for projection onto
$\mathbb{PP}_d$~\cite{draper_2013_admm_lp}. Follow-up work by X.~Zhang
and Siegel provided a new approach based on identifying the facet of
the polytope to be projected
onto~\cite{siegel_2013_projection_lp}. Additionally, G.~Zhang et
al. made improvements by reducing the parity polytope projection to a
projection onto the probability
simplex~\cite{kleijn_2013_pp_projection}. In a previous work, we
extracted the best features of these new algorithms and proved their
compatibility~\cite{wasson_2015_pp}. This results in a new projection
method appropriate for hardware. Our approach uses a parallizable
method for efficient projection onto the $d$-dimensional probability
simplex~\cite{duchi_2008_simplex_projection}.

Fig.~\ref{ppProject} displays the geometric interpretation of the parity polytope projection algorithm developed in~\cite{wasson_2015_pp}. First, the facet of the polytope on which the projection lies is identified with the cut-search algorithm of \cite{siegel_2012_cut_search}. Then, a similarity transform, derived from that developed in \cite{kleijn_2013_pp_projection}, is performed using this information. This reduces the parity polytope projection to a projection onto the probability simplex. After the simplex projection, the transform is inverted to obtain the final projection onto the parity polytope.

Our previous study revealed how to achieve a fully parallel hardware implementation of the projection algorithm using primitives like sorting networks and prefix sum operations~\cite{wasson_2015_pp}. The hardware implementation achieves an area scaling on the order of $\mathcal{O}\left(d \left(\log d\right)^2\right)$ with a delay scaling of $\mathcal{O}\left( \left(\log d\right)^2\right)$, where $d$ is the projection dimension.

\begin{figure*}
	\centering
	\includegraphics[width=0.89\textwidth]{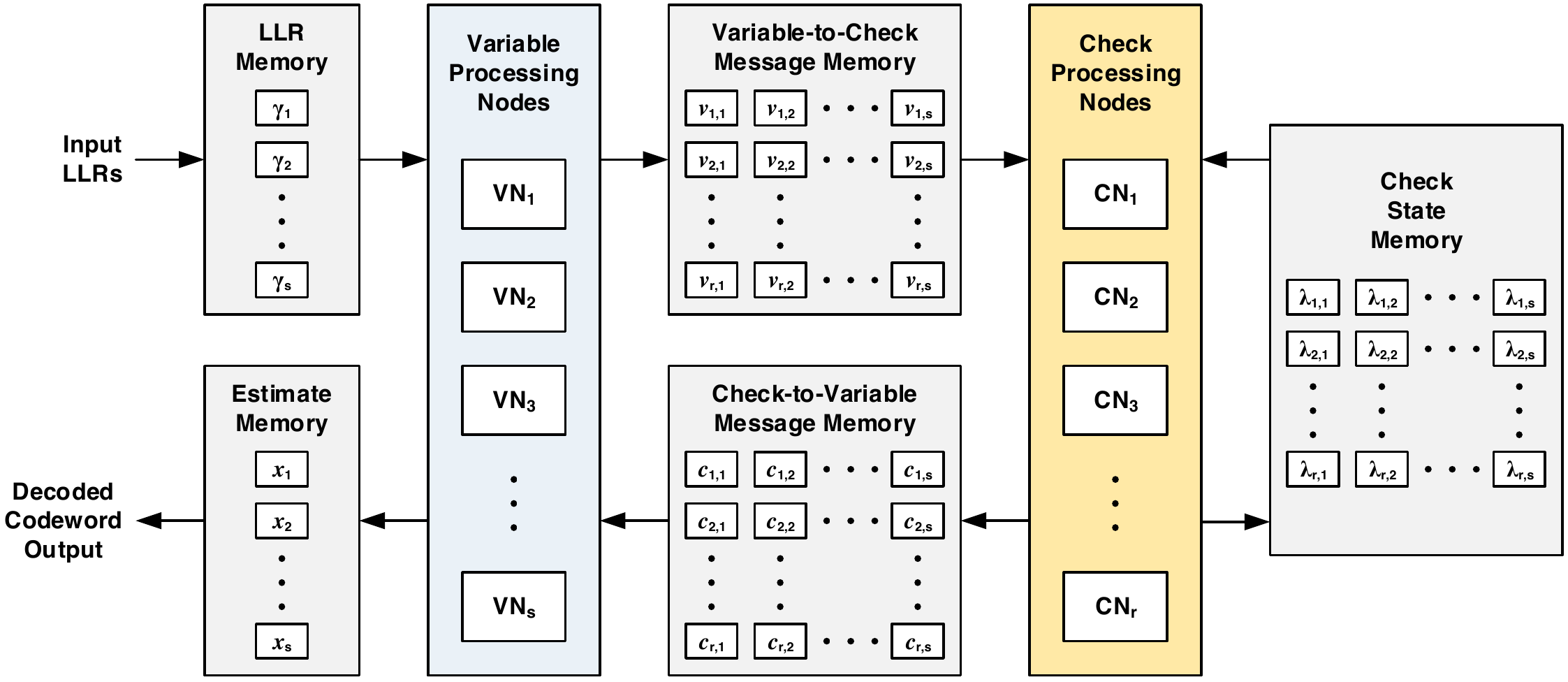}
	\caption{Partially-parallel decoder architecture.}
	\label{decoderArchitecture}
\end{figure*}
\section{Hardware Architecture}

In the previous section, we showed that ADMM-LP decoders implement a message-passing schedule similar to BP decoders, one that is based on the implied check and variable node connectivity structure defined by the binary parity-check matrix. This similarity allows us to build upon  well-known hardware architectures used for BP decoders in the design of a hardware-based ADMM-LP decoder implementation.  We modify the arithmetic kernels in the check and variable processing nodes as per the operations outlined in Algorithm~\ref{decodingALgorithm}. 

\subsection{Architecture Selection for FPGA Platform}

Hardware architectures for BP decoders have been studied extensively over the past 15 years, and can be classified into one of the following three architectures: fully-parallel, partially-parallel, and serial. Fully-parallel decoders achieve high data throughput at the expense of high area/computational resource utilization, while serial decoders require minimal area but suffer from high latency \cite{Blanksby2002, Bhatt2006}. Partially-parallel decoders provide the most optimal trade-off between area resource utilization and data throughput \cite{Hocevar2004}, and are generally well-suited for FPGA implementation since power consumption is not a crucial metric for this work. 

The goal of this work is to develop a platform enabling accelerated ADMM-LP decoding in order to study the error correction performance of a wide variety of linear binary codes. An FPGA-based platform provides a re-programmable and cost-effective solution. Once a code's performance is well-understood, a custom, energy-efficient architecture can later be explored for silicon-based integrated circuit implementation. In this work, however, we sacrifice the power, area, and throughput design objectives to create a more general implementation that can rapidly be applied to study new codes. This work therefore implements a partially-parallel decoder architecture, which allows us to take advantage of the high availability of FPGA slice registers for deep pipelining to minimize decoding latency, while operating within the fixed logic and routing resource limitations of the target FPGA. 

\subsection{Partially-Parallel Decoder Implementation}

A central challenge in implementing hardware-based decoders is the scalability of the message-passing network, which often requires resource-intensive wiring and memory interconnect resources to pass messages between check node (CN) and variable node (VN) processing units. The partially-parallel architecture allows us to minimize FPGA routing complexity by implementing the message-passing network with regularly-distributed, on-chip FPGA block RAMs. Fig.~\ref{decoderArchitecture} presents an overview of our partially-parallel QC-LDPC decoder architecture.  The architecture  is comprised of multiple memory types to store input LLRs, intermediate messages, and output codewords, as well as pipelined CN and VN processing units that perform the arithmetic operations used in Algorithm~\ref{decodingALgorithm}. 

We restrict ourselves to quasi-cyclic (QC) codes~\cite{kou_2001_qc,fossorier_2004_qc} in order to  simplify message routing and memory interfacing. QC codes are defined by a parity-check matrix formed by tilings of $p\times p$ circulant matrices. Therefore, each tile of a QC parity-check matrix can either be the all-zeros matrix or some addition of shifted-identity matrices. The tilings naturally divide the parity-check matrix into $s:=\frac{n}{p}$ ``proto''-columns and $r:=\frac{m}{p}$ proto-rows. Inside a given proto-row (column), the required message locations for a check (variable) computation are the locations for the previous check (variable) plus 1 modulo $p$. This rich class of codes is popular in hardware implementations, appearing in standards such as IEEE 802.11ad (WiGig)~\cite{IEEE-80211ad}.

The first execution step our decoder performs is to load channel LLRs into memory. We instantiate $s$ memories, each of depth $p$ to store the LLRs. Each of these memories is then read in parallel to feed LLRs into $s$ pipelined VNs. The VNs also receive messages from a check-to-variable (CN-to-VN) message memory, to be discussed later. At the output of the VNs, the current variable estimates (the $x_i$'s) are written into $s$ estimate memories in parallel, to be read from upon decoding termination. Additionally, variable estimates are written into variable-to-check (VN-to-CN) message memories. There is a VN-to-CN message memory for each shifted-identity matrix used to construct the parity-check matrix. These memories are addressed using their corresponding shift number to ensure the messages are passed to the proper CN. 

Next, $r$ pipelined CNs read their required messages in parallel from the VN-to-CN message memory. Additionally, the check states are read from check state memories, which are instantiated in the same manner as the VN-to-CN message memories. However, address shifting is not required since these memories are only written to, and read by, CNs. When a CN computation completes, the new check states are written back into the check state memory and the messages are written into CN-to-VN message memories. These message memories are again structured in the same manner with write operations using cyclic shift information. The process  repeats until the maximum number of iterations is exceeded, or some early termination condition is satisfied.

\subsection{Fixed-Point Message Quantization}

In our current implementation, we have found the ADMM-LP decoder to be
sensitive to fixed-point quantization.  In contrast to BP decoders,
which can be implemented with 5 or 6-bit variable widths with minimal
degradation in bit-error-rate performance compared to floating point
\cite{Park2014}, ADMM requires larger bit-widths. We believe that the
ADMM-LP decoder requires higher precision because the result of the
projection operation that check nodes perform must be quantized.  This
results in a loss of precision and a corresponding deterioration of
message resolution.

We now discuss some intuition behind the choices we made in picking
fixed-point representations.  We first note that a change in the
assignment of bits between integer and fraction parts of fixed-point
LLRs amounts to a linear scaling of the objective. However, any
scaling of the objective in an LP (i.e., of $\gamma$
in~(\ref{LPDecoding})) does not change the solution of the LP.  This
provides some flexibility in choosing the fixed-point representation
of the LLRs.  Next we note that each message passed to a variable node
can be thought of as either trying to overcome the channel information
or as trying to reinforce it. Thus, any extra bit-width should be
allocated to the integer part of a CN-to-VN message. This provides
dynamic range to override channel LLRs.  In contrast, any extra bit-width allocated to VN-to-CN messages should be fraction bits.  An
increase in the number of fraction bits mitigates the effect of the
inexact (due to finite precision) normalization by
$|\mathcal{N}_{v\left( i \right)}|$ in the variable nodes
(cf.~step~\ref{step.varUpdate} of the Algorithm).

Based on the above design intuition, we select fixed-point message
representations based on the premise of retaining as much channel
information as possible. First, we consider the bit-width for both the
LLRs and the estimate outputs.  These, respectively, correspond to the
decoder's input and output message widths. Next, we consider how many
additional bits VN-to-CN and CN-to-VN messages will receive. VN-to-CN
messages, as well as the estimates, lie in the unit
hyper-cube. Therefore, these messages receive one sign bit, one integer
bit, and allocate the remainder to fraction bits. 
Next, we give LLRs one sign bit, zero integer bits, and allocate the remainder
to fraction bits. This ensures that all channel information is visible in
the estimates and the VN-to-CN messages. The CN-to-VN messages are
given one sign bit and the same number of fraction bits as the
LLRs. This is done so that the summation in the VN computation
produces an output that does not have any constant bits for some given
LLR. Finally, the check states are given the same representation as
the CN-to-VN messages because they are computed in a similar manner.

The next section presents the error-correction performance results for
our FPGA-based ADMM-LP decoder for two different linear block codes.
We also explore the resources required by this architecture on a
state-of-the-art FPGA.

\section{Results}

\subsection{Performance}
In order to test the hardware viability of ADMM-LP decoding, we
developed an FPGA-in-the-loop simulation environment using an Xilinx
Virtex 5 FPGA. The proposed architecture was synthesized for the FPGA
along with logic for random number generation and data transfer.

The binary-input AWGN channel was simulated using a synthesized Gaussian random
number generator~\cite{liu_2015_rng}. The core is a linear feedback
shift register with period $2^{176}$ fed into an inverse cumulative
distribution function approximation. The output of the simulated
channel was saturated at one standard deviation of channel noise to
create LLRs within the decoder's input range. This on-FPGA method of
channel simulation was necessary since generating channel outputs on a
PC and transferring them to the FPGA became a simulation
bottleneck. We verified that the on-FPGA method produced equivalent
channel simulations for the decoder configuration we present.

\begin{figure}[h]
	\centering
	\includegraphics[width=0.45\textwidth]{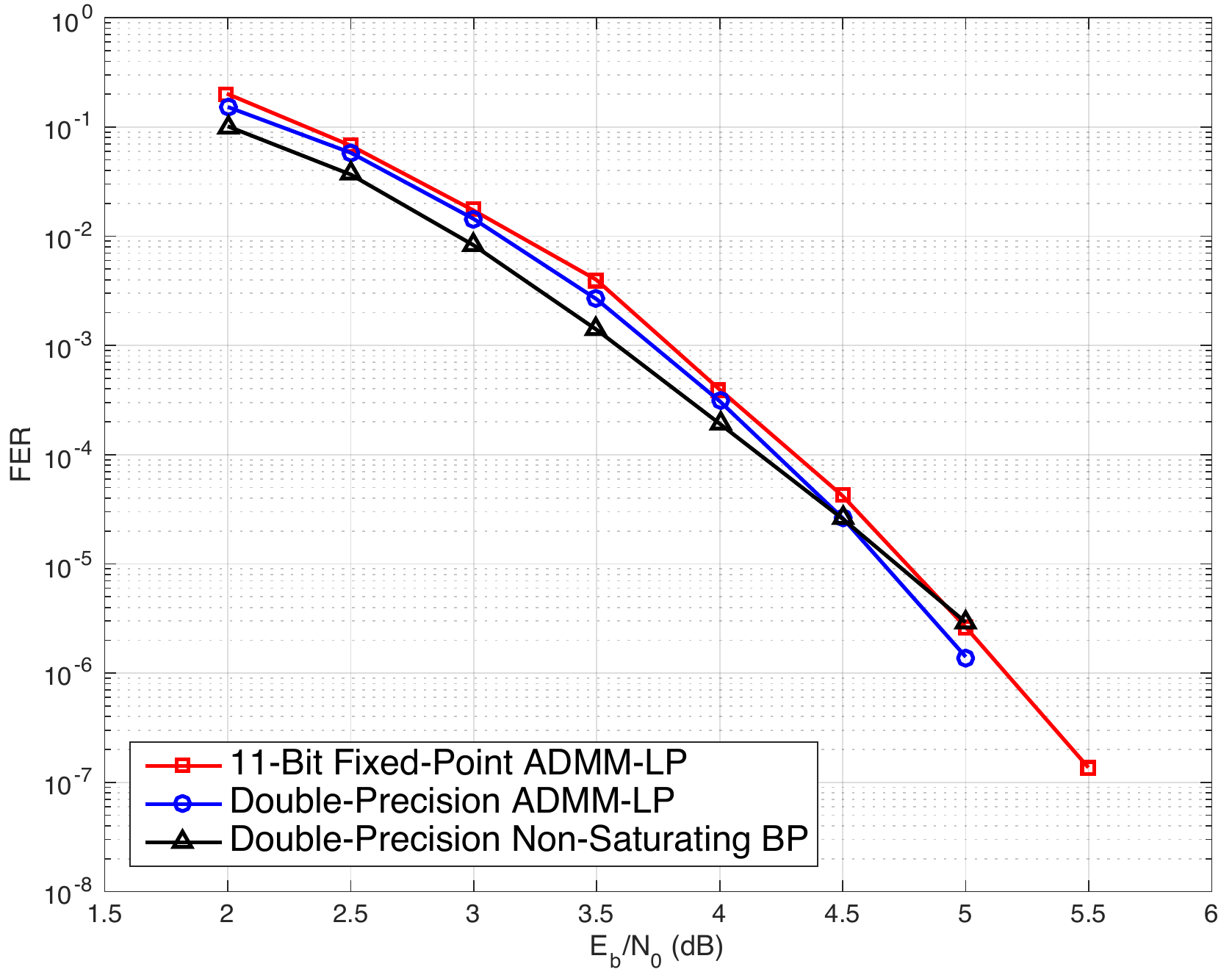}
	\caption{FER performance of the $[155,64]$ ``Tanner'' code.}
	\label{tannerCode}
\end{figure}

\begin{figure}[h]
	\centering
	\includegraphics[width=0.45\textwidth]{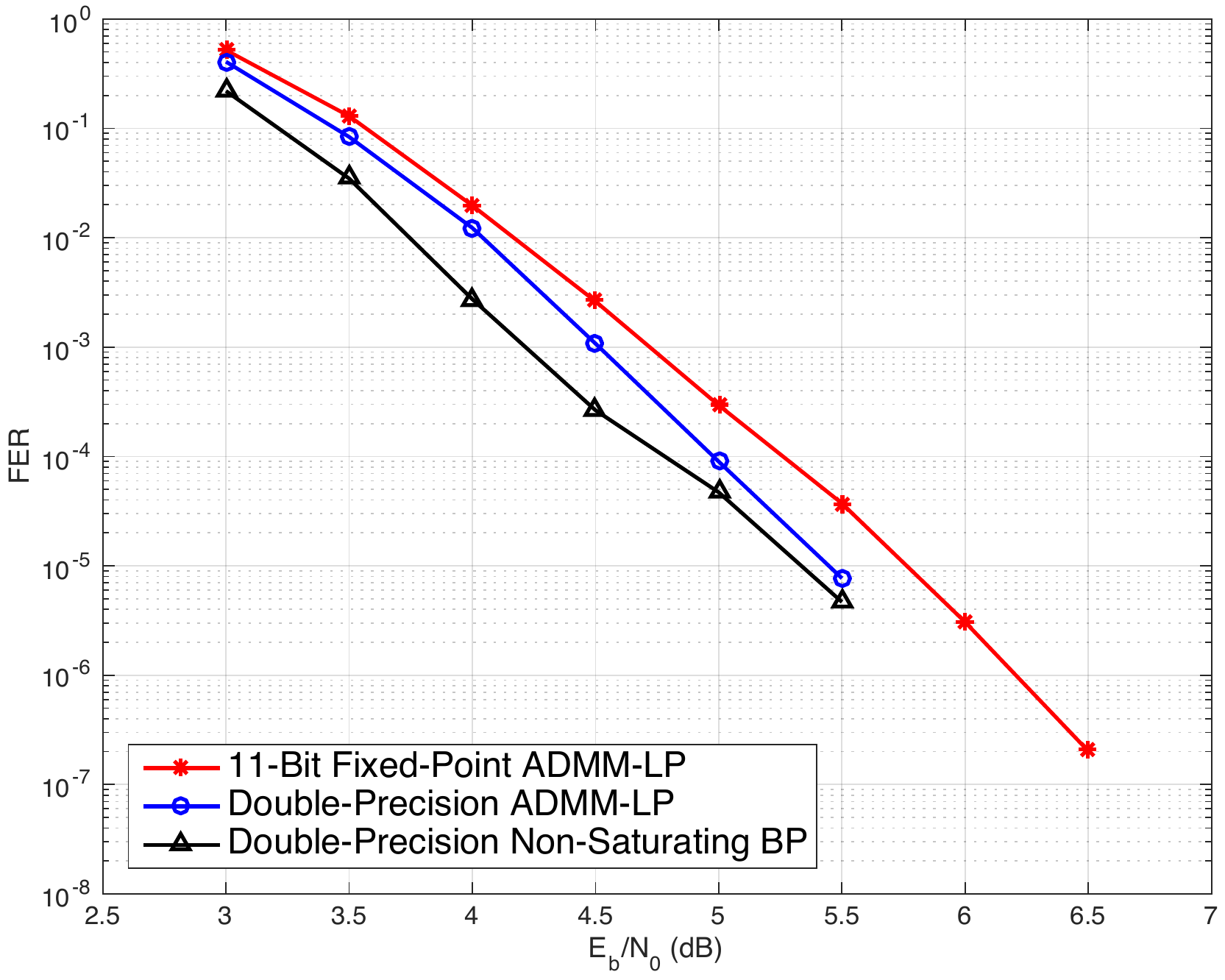}
	\caption{FER performance of the WiGig code.}
	\label{wifiCode}
\end{figure}

Both the aforementioned $n=155$, rate $64/155$ code given by Tanner
and the $n=672$, rate $13/16$ WiGig code were used to test the
fixed-point ADMM-LP hardware implementation. We configured our
implementation to receive 8-bit LLRs and pass 11-bit messages
internally. Both codes were also simulated using double-precision
ADMM-LP and non-saturating sum-product BP
decoding~\cite{barmanSiegel_2014_nonSat}. All implementations were run
for a maximum of 500 iterations and each point shown on the plots is
an accumulation of at least 100 frame errors.

Fig.~\ref{tannerCode} shows that the fixed-point implementation of
ADMM-LP decoding produces error rates extremely close to
double-precision implementations, even surpassing double-precision BP
at a high SNR. Fig.~\ref{wifiCode} displays similar competitive
error rates for the WiGig code.

One observed effect of the fixed-point implementation we noted is a
slight codeword asymmetry. Specifically, for small bit-widths, the
all-zeros codeword achieved a lower FER than higher-weight codewords. The effect is due to numerical truncation rounding toward negative infinity. Further investigation, not shown in this paper, suggests that including rounding operations as well as centering estimates about zero resolves the codeword asymmetry problem. However, due to this effect, we use high-weight codewords in the displayed simulations.

\subsection{Resource Utilization}

We now detail the resources required to synthesize the decoder on a
state-of-the-art FPGA. The FPGA used in the following resource
estimates is an Altera Stratix V FPGA (model 5SGXEA7N2F45C2). This
FPGA has 234,720 adaptive logic modules (ALMs), 256 dedicated DSP
blocks, and 2,560 M20K RAM blocks.

First we consider synthesizing the decoder architecture for the Tanner
code. This is a regular QC-LDPC code with degree-5 check and degree-3
variable nodes. The parity-check matrix is composed of three
proto-rows and five proto-columns. Therefore the decoder is composed
of five degree-3 variable nodes and three degree-5 check nodes.

The WiGig code is also quasi-cyclic. However, some tiles of the
parity-check matrix are all-zeros. Therefore this code is
irregular. This results in an implementation with fourteen degree-3 variable
nodes, one degree-2 variable node, and one degree-1 variable
node. Additionally, there is one degree-16 check node, one degree-15
check node, and one degree-14 check node.

\begin{table}[h]
	\centering
	\begin{tabular}{| c | c  | c | c | c | c | }
		\hline		
		Module & 
		\begin{tabular}{@{}c@{}}ALM  \\ (\%)\end{tabular} &
		\begin{tabular}{@{}c@{}}DSP  \\ (\%)\end{tabular} &
		\begin{tabular}{@{}c@{}}RAM  \\ (\%)\end{tabular} &
		\begin{tabular}{@{}c@{}}Period  \\ (ns)\end{tabular} &
		 \begin{tabular}{@{}c@{}}Pipeline \\ Stages\end{tabular}  \\ \hline\hline
		{\bf Tanner Dec.} & {\bf 5.28} & {\bf 4.30} & {\bf 1.64} & {\bf 4.72} & {\bf -}  \\
		Deg. 3 VN & 0.06 & 0.39 & 0 & 4.72 & 10  \\ 
		Deg. 5 CN & 1.48 & 0.78 & 0 & 4.72 & 46  \\ 
		\hline \hline 
		{\bf WiGig Dec.}& {\bf 15.06} & {\bf 17.97} & {\bf 3.59} & {\bf 4.58} & {\bf -}  \\
		Deg. 1 VN & 0.02 & 0 & 0 & 4.58 & 9  \\ 
		Deg. 2 VN & 0.03 & 0 & 0 & 4.58 & 10  \\ 
		Deg. 3 VN & 0.03 & 0.39 & 0 & 4.58 & 10  \\ 
		Deg. 14 CN & 4.37 & 3.91 & 0 & 4.58 & 53  \\ 
		Deg. 15 CN & 4.67 & 4.30 & 0 & 4.58 & 53  \\ 
		Deg. 16 CN & 5.11 & 4.30 & 0 & 4.58 & 54  \\ 
		\hline 
	\end{tabular}
	\caption{Resource utilization table for Altera Stratix V.}
	\label{resourceTable}
\end{table}

Table~\ref{resourceTable} displays the percentage of on-FPGA resources
required to synthesize the decoders for each of the two codes (cf.~the
bolded rows labeled ``Tanner Dec.'' and ``WiGig Dec.'').  We also give
numbers for each of the important sub-modules. The clock period of the
synthesized circuit (which is constant for all modules in a given
decoder), and the number of pipeline stages for sub-modules, are also
provided.

\section{Conclusion}
This work presented an early investigation into the feasibility of a
hardware-based ADMM-LP decoder.  We target an FPGA platform with fixed
logic, memory, and routing resources. We showed that an FPGA-based,
partially-parallel, decoder architecture can be used to study ADMM-LP
decoding performance of linear block codes shorter than 1000 bits.  We
now mention some future work.  The first is the full
understanding (and elimination) of the codeword asymmetry
mentioned. The second is further investigation of error-floor regime
performance for LP (and penalized-LP~\cite{liu_2016_penalized})
decoding.  A third is the development of simplified decoding
algorithms that maintain error-floor performance while reducing the
required bit-width.  Our ultimate objective is a fully-custom silicon
integrated circuit implementation, which would be
required to achieve decoding speedup for longer codes. In such an
implementation, it would be beneficial to explore new hardware
architectures that would provide greater information throughput, and
parity-check matrix reconfigurability.

\bibliographystyle{IEEEtran}
\bibliography{IEEEabrv,sipsLP}
	
\end{document}